\documentclass[aps,showpacs,showkeys,preprint,amsmath,amssymb]{revtex4}

\usepackage{natbib}  
\bibpunct{(}{)}{,}{a}{}{,} 

\usepackage{graphicx}
\usepackage{dcolumn}
\usepackage{bm}

\begin{document}

\title{Differential Renormalization-Group Approach 
\\ to the Layered sine-Gordon Model}
\author{I. N\'andori}
\email{nandori@atomki.hu}
\affiliation{Institute of Nuclear Research of the Hungarian Academy 
of Sciences, H-4001 Debrecen, P.O.Box 51 Hungary; \\
Max Planck Institute for Nuclear Physics, P.O.Box 103980, 
Heidelberg 69029, Germany
}%
\author{K. Sailer}
\affiliation{ Department of Theoretical Physics, University of Debrecen,
H-4010 Debrecen, P.O.Box 5 Hungary
}%

\date{\today}

\begin{abstract}

New qualitative picture of vortex length-scale dependence has been 
found in recent electrical transport measurements performed on 
strongly anisotropic $\rm{Bi_2 Sr_2 Ca Cu_2 O_8}$ single crystals in 
zero magnetic field. This indicates the need for a better description 
of the 3D/2D crossover in vortex dimensionality. The vortex-dominated
properties of high transition temperature superconductors with 
extremely high anisotropy (layered systems) are reasonably well 
described in the framework of the layered XY model which can be mapped
onto the layered sine-Gordon model. For the latter we derive an 
exact renormalization group (RG) equation using Wegner's 
and Houghton's approach in the local potential approximation.
The agreement of the UV scaling laws find by us 
by linearizing the RG equations with those obtained previously in the 
literature in the dilute gas approximation makes the improvement 
appearant which can be achieved by solving our RG equations numerically.  
\end{abstract}

\pacs{Valid PACS appear here}
\keywords{Renormalization group method, layered sine-Gordon model}

\maketitle

\section{\label{intro} Introduction}
Vortex dominated properties of high-temperature superconductors
(HTSC) are usually investigated by means of electrical transport 
measurements which have raised many questions about the critical 
behaviour and the length-scale dependence of vortices in layered 
systems. The HTSC materials have a layered structure and usually 
consist of copper-oxide superconducting planes separated by insulators. 
In case of extremely high anisotropy like in $\rm{Bi_2 Sr_2 Ca Cu_2 O_8}$
(BSCCO) single crystals, the coupling between the $\rm Cu O_2$ 
planes is weak. Due to this weak coupling, one of the commonly accepted 
models is the layered XY model \citep{pierson_lsg} where the weak 
interlayer 
coupling is given by a Lawrence-Doniach type term \citep{pierson_rg}. 
The topological excitations in superconducting layers are 
vortex-antivortex pairs. Two such pairs belonging to neighbouring 
layers can form vortex loops due to the Josephson coupling.

The layered XY model can be mapped onto the layered sine-Gordon (LSG) 
model \citep{pierson_map,pierson_lsg} or it can be transformed into the 
layered Coulomb gas (LCG) model \citep{pierson}. 
The critical behaviour of vortices and vortex loops in the LSG and LCG 
models have been investigated by renormalization group (RG) methods 
using the dilute gas approximation \citep{pierson_rg,pierson}   
which assumes low vortex fugacity. However, the vortex fugacity becomes 
large above the critical temperature (but below the Ginzburg-Landau 
transition temperature) when individual vortices can be spontaneously 
created although the sample is still in the superconducting phase.
  
Recent current transport measurements \citep{kalman} performed on 
BSCCO superconducting single crystals show a new qualitative picture of 
vortex behaviour above the critical temperature \citep{csmag}. Using a 
multi-contact 'DC flux transformer' configuration, the temperature 
dependence of primary and secondary voltages  are measured in zero 
magnetic field. (The side of the crystal where the current is injected  
is called the primary one, its opposite side is the secondary one.)
According to the theoretical predictions based on the dilute gas 
approximation \citep{pierson}, above the critical temperature one 
should expect the decrease of the secondary voltage to zero as a 
function of the temperature. Instead, the experimental data reveal 
a new feature \citep{csmag}, a second peak appears in the temperature 
dependence of the secondary voltage. A possible reason for the 
disagreement between the theory and experiment is the usage of the 
dilute gas approximation. This indicates the need for a better 
description of length-scale dependence of the vortex dimensionality 
at the 3D/2D crossover and hence the modification of the RG analysis 
\citep{csmag}.
 
The purpose of this paper is to propose a theoretical study of the 
LSG model using the differential RG approach \citep{wilson1,wilson2} 
in order to improve the treatment of the length-scale dependence of 
vortices. In section 2, the definition of the layered models and their 
interrelations are summarized. 
After giving a short sketch of Wegner's and Houghton's RG method 
\citep{wh} for models with two interacting scalar fields in section 3,
we apply it to the LSG model using the local potential approximation 
\citep{janosRG} in section 4. The UV scaling laws are found and 
discussed in section 5 on the base of the linearized RG equations.
The comparison of the UV scaling laws to those obtained previously 
in Refs. \citep{pierson_lsg,pierson_rg} using a momentum-space RG 
method with  smooth cut-off in the dilute gas approximation is also 
given, and the possibility of an improved treatment of the RG flow    
by our method is pointed out. Finally the results are summarized in 
section 6.

\section{\label{lsgsect} Layered Models}
The phase structure of HTSC materials is usually investigated in 
the framework of the Ginzburg-Landau (GL) theory of superconductivity 
at the phenomenological level. The GL theory is based on a variational 
method which has been applied to the gradient expansion of the 
free-energy in powers of the complex order parameter $\psi$ and 
its derivatives $\nabla\psi$ where $\vert\psi\vert^2$ represents 
the local density of superconducting electron pairs. In case of 
strong anisotropy and in the absence of external fields  the
free energy becomes \citep{csmag}
\begin{eqnarray}
F = \sum_n \int {\rm d}^2 r \, \left({1\over 2} J_{\parallel}
(\nabla\phi_n)^2 + J_{\perp} [1-\cos(\phi_n - \phi_{n-1})]\right),
\end{eqnarray}
where $\psi_n(r) = \psi_0 \exp[i \phi_{n}(r)]$ is assumed with
$\psi_0$  real and identical in every layer $n$. The coupling 
constants of the model are $J_{\parallel}= {\hbar^2 \psi_0^2/ m_{ab}}$ 
and $J_{\perp}= {\hbar^2 \psi_0^2/ m_{c} s^2}$ with the effective 
masses $m_{ab}, m_{c}$, and the interlayer distance $s$. 
The last term is related to the Lawrence-Doniach term, in which the 
coupling between the layers is the Josephson coupling. Expanding the 
cosine term in Taylor series around zero, one arrives at the layered 
two-dimensional XY model \citep{pierson_lsg}:
\begin{eqnarray}
H_{LXY} = {1\over 2} \sum_n \int {\rm d}^2 r \, J_{\parallel}
(\nabla\phi_n)^2 + {1\over 2} \sum_n \int {\rm d}^2 r \, J_{\perp}
(\phi_n - \phi_{n-1})^2 .
\end{eqnarray}
The layered XY model can be mapped onto the LSG or LCG models 
\citep{pierson_map,pierson_rg}. For a system consisting of two 
layers the Lagrangian of the LSG model reads as \citep{pierson_lsg}
\begin{eqnarray}\label{lsg}
L_{LSG} = {1\over 2} [(\partial \phi_1)^2 + (\partial \phi_2)^2]
+   {1\over 2} {\tilde J} (\phi_1 -\phi_2)^2 
+ {\tilde u} \left[\cos(\beta\phi_1) + \cos(\beta\phi_2)\right],
\end{eqnarray}
where $\phi_1$, $\phi_2$ are one-component scalar fields, 
$\tilde u$ is the fugacity parameter for the vortices of the model,  
$\beta^2 = J_{\parallel}$ and 
$\tilde J= J_{\perp}/(\Lambda^2 J_{\parallel})$ are dimensionless 
coupling constants and $\Lambda$ stands for the UV momentum cutoff. 
The second term on the r.h.s. describes the interaction of the layers.
In Refs. \citep{pierson_lsg,pierson_rg} RG flow equations,
\begin{equation}\label{dilute}
k {d\over dk} \beta^2 = {1\over 4\pi} \left({\tilde u}^2 \beta^4 - 
4.5 \beta^4 {\tilde u}^2 {\tilde J} \ln{\tilde J}\right), 
\hspace{0.5cm}
k {d\over dk} {\tilde u}
 = {\tilde u} 
\left({\beta^2\over 4 \pi} -2 \right),
\hspace{0.5cm}
k {d\over dk} {\tilde J} = - 2 {\tilde J} ,
\end{equation}
where $k$ is the momentum scale,
have been obtained
 for the coupling 
constants of the LSG model (\ref{lsg}) with the help of a momentum 
space RG analysis using a smooth  (anisotropic) momentum cutoff
and the dilute gas approximation. The latter assumes that the 
fugacity of the vortex gas, i.e. the coupling constant $\tilde u$ 
in (\ref{lsg}) is small.

\section{\label{rg} Differential RG approach}
The renormalization of the LSG model is presented in this section
by means of the differential RG approach performed in momentum space
with sharp cutoff \citep{janosRG}. The blocking transformations 
\citep{wilson1,wilson2} are realized by successive elimination of the 
field fluctuations according to their decreasing momentum in 
infinitesimal steps. We use Wegner's and Houghton's method (WH-RG) 
\citep{wh} when the high-frequency modes are integrated out above the 
moving sharp momentum cutoff $k$ and the physical effects of the 
eliminated modes are built in the scale-dependence of the coupling 
constants. 

Assuming that the generating functional does not change under 
the infinitesimal blocking transformation when the momenta of the 
eliminated modes are taken from a thin momentum shell with radius 
$k$ and thickness $\Delta k$, the blocking relation  
\begin{equation}
e^{-S_{k-\Delta k}[\phi]}=\int{\cal D}[\phi']e^{-S_k[\phi+\phi']},
\label{WHblocking}
\end{equation}
is obtained for the blocked action $S_k[\phi]$ \citep{janosRG}. 
The field variables $\phi$ and $\phi'$ contain Fourier modes  
with momenta $p<k-\Delta k$ and $k-\Delta k < p < k$, respectively. 
Expanding the action $S_k[\phi+\phi']$ in Taylor series around its 
saddle point, the path integral can be evaluated and the WH-RG method 
\citep{wh} provides the functional RG equation 
\begin{equation}\label{rgeq}
S_{k-\Delta k}[\phi] =  S_{k}[\phi] + {\hbar\over 2} {tr} 
\ln\left(S''_k[\phi]\right) + {O}(\hbar^2) , 
\end{equation}
 where $S''_k$ denotes the
second functional derivative of the blocked action with respect to 
the field and the trace is over the momentum shell [$k-\Delta k,k$]. 
In the limit $\Delta k\to 0$ Eq. (\ref{rgeq}) reduces to an exact 
integro-differential equation for the functional $S_k[\phi]$
because the small parameter $\Delta k/k$ suppresses the higher-loop 
contributions \citep{janosRG}. The generalization of Eq.(\ref{rgeq})  
for two, interacting scalar fields is straightforward following the 
method proposed in \citep{n_component} and results in  
\begin{equation}\label{WHfunc} 
S_{k-\Delta k}[\phi_1, \phi_2] = S_k[\phi_1,\phi_2] + {\hbar\over 2} 
{tr} \ln\left(S^{11}_k[\phi_1, \phi_2] S^{22}_k[\phi_1, \phi_2] - 
S^{12}_k[\phi_1, \phi_2] S^{21}_k[\phi_1, \phi_2]\right),  
\end{equation}
where $S^{ij}_k[\phi_1,\phi_2]$ denote the second functional 
derivatives of the blocked action with respect to $\phi_i$ and 
$\phi_j$.

If the saddle point is non-trivial in (\ref{WHblocking}), a spinodal 
instability occurs at some critical momentum scale  $k_c$, and the 
WH-RG equation (\ref{rgeq}) loses its validity \citep{tree}. 
Then, the blocking produces a  tree-level effect \citep{tree}. 
In case of two interacting scalar fields the tree-level blocking 
relation reads as
\begin{equation}\label{treefunc}
S_{k-\Delta k} [\phi_1, \phi_2] = \min_{\phi^{cl}_1, \phi^{cl}_2} 
\left(S_k[\phi_1 + \phi^{cl}_1, \phi_2 + \phi^{cl}_2] \right),
\end{equation}
where the minimum is sought over field configurations including 
only Fourier-modes with momenta from the infinitesimal momentum 
shell $k-\Delta k <p<k$.

\section{\label{rg-lsg} RG approach to the generalized LSG model}
In this section we specify the WH-RG equation (\ref{WHfunc}) and the 
tree-level blocking relation (\ref{treefunc}) for the LSG model.
First, the WH-RG equation and the tree-level blocking relation, 
have to be projected into a particular functional subspace, in order 
to reduce the search for a functional (the blocked action) to that 
of functions. Here we  assume that the blocked action contains only 
local interactions. Expanding it in powers of the gradients of the 
fields $\phi_1$ and $ \phi_2$ and keeping only the leading order 
terms one arrives at the local potential approximation (LPA) 
\citep{janosRG}. In LPA the ansatz for the blocked action for two 
scalar fields can be written as
\begin{equation}
S_k [\phi_1,\phi_2] = \, \int {\mathrm d}^d x \left[
 {1\over2} \, (\partial_{\mu}\phi_1)^2 + {1\over2} 
(\partial_{\mu}\phi_2)^2 \, + \, V_k (\phi_1,\phi_2) \right]
\end{equation}
with the blocked potential $V_k (\phi_1,\phi_2)$ satisfying the WH-RG 
equation 
\begin{equation}
\label{WH}
k\partial_k V_k(\phi_1,\phi_2) = - k^d \alpha_d
\ln \left({[k^2 + V^{11}_k] [k^2 + V^{22}_k] -
[V^{12}_k] [V^{21}_k] \over  k^4} \right)
\end{equation}
with $\alpha_d = \frac{1}{2} \Omega_d (2\pi)^{-d}$, where $\Omega_d$
denotes the solid angle in $d$ dimensions and $V^{ij}_k= \partial_{\phi_i}
\partial_{\phi_j}V_k (\phi_1,\phi_2) $.
 In order
to remove the trivial scale-dependence of the coupling constants, the 
WH-RG equation (\ref{WH}) has to be rewritten in terms of dimensionless
quantities. For dimensions $d=2$, the dimensionless form of Eq.(\ref{WH}) 
reads as 
\begin{equation}
\label{WHdim}
\left(2 + k\partial_k \right) {\tilde V_k(\phi_1,\phi_2)} =
\,-  \, \alpha_2 \, \ln\left(
[1 + {\tilde V^{11}_k }] [1 + {\tilde V^{22}_k }] 
-  {\tilde V^{12}_k }  {\tilde V^{21}_k } \right)\,,
\end{equation}
where  ${\tilde V_k(\phi_1,\phi_2)} = k^{-2} V_k(\phi_1,\phi_2)$ has 
been introduced.

The argument of the logarithm in Eqs. (\ref{WH}) and
(\ref{WHdim}) must be positive. If the argument vanishes or if it
changes sign at a critical value $k_{\mathrm c}$, the WH equation 
(\ref{WH}) loses its validity for $k<k_{\mathrm c}$. This is a 
consequence of the spinodal instability, i.e. that of the vanishing 
of the restoring force for field fluctuations of momenta from the 
infinitesimal momentum shell. Such fluctuations can grow to those 
of finite amplitude and result in a non-vanishing sadlle-point 
configuration. Below the critical scale $k<k_c$ the tree-level 
blocking relation (\ref{treefunc}) can be used to find the saddle 
point. Looking for it among the configurations $\phi^{cl}_1$ and 
$\phi^{cl}_2$ having the form of plane waves one finds for the 
blocked local potential
\begin{equation}\label{treenew}
V_{k-\Delta k}(\phi_1,\phi_2)
= \min_{\rho_k} \left[  2 \rho^2_k k^2 + 
{1\over2} \int_{-1}^{1} {\mathrm d}p  
V_{k}(\phi_1+2\rho_k \cos(\pi p),\phi_2+2\rho_k \cos(\pi p)) \right],
\end{equation}
where $\rho_k$ is the amplitude of the plane waves \citep{tree}.
Notice that the layers in the LSG model are physically equivalent 
therefore the potential is invariant under the  transformation 
$\phi_1 \leftrightarrow \phi_2$ which implies identical plane-wave 
forms of   $\phi^{cl}_1$ and $\phi^{cl}_2$ with the same amplitude 
$\rho_k$ and wave vector in the saddle point. For dimensions $d=2$, 
the tree-level relation (\ref{treenew}) for dimensionless quantities 
reads as 
\begin{equation}\label{treenew2}
{\tilde V}_{k-\Delta k}(\phi_1,\phi_2)
= \min_{\rho_k} \left[  2 \rho^2_k  + 
{1\over2} \int_{-1}^{1} {\mathrm d}p  
{\tilde V}_{k}(\phi_1+2\rho_k \cos(\pi p),\phi_2+2\rho_k \cos(\pi p)) 
\right].
\end{equation}

As to the next, we specify the WH-RG equation (\ref{WHdim}) and the 
tree-level blocking relation (\ref{treenew2}) for the LSG model. On 
the one hand, the ansatz for the blocked potential should be reach 
enough in order to ensure that the RG flow does not leave the chosen 
subspace of blocked potentials. On the other hand, the ansatz for the 
blocked potential should preserve all symmetries of the original 
model at the UV cutoff scale $k=\Lambda$. 
In particular, the blocked potential for the LSG model should 
be invariant under the exchange of the field variables, 
$\phi_1 \leftrightarrow \phi_2$ because the layers are physically 
equivalent and also reveal the symmetries $\phi_i \to -\phi_i$ and 
$\phi_i \to \phi_i + {2\pi\over \beta^2}$ (c.f. Eq. (\ref{lsg})).
Therefore, we choose the ansatz 
\begin{equation}\label{def1}
{\tilde V}_{k}(\phi_1,\phi_2) = {1\over2} {\tilde J_k} 
(\phi_1 - \phi_2)^2 +  {\tilde U}_{k}(\phi_1,\phi_2),
\end{equation}
where ${\tilde U}_{k}(\phi_1,\phi_2)$ is an arbitrary periodic 
function of the fields 
\begin{equation}\label{def2}  
{\tilde U}_{k}(\phi_1,\phi_2) = 
\sum_{n,m} \big[ {\tilde u}_{nm}(k) 
\cos(n \beta \phi_1)\cos(m \beta \phi_2) 
+ {\tilde v}_{nm}(k) \sin(n \beta \phi_1) \sin(m \beta \phi_2) \big]
\end{equation}
with symmetric Fourier amplitudes  ${\tilde u_{nm}} = {\tilde u_{mn}}$, 
${\tilde v_{nm}} = {\tilde v_{mn}}$. The potential (\ref{def2}) 
defines the generalized LSG model which contains higher harmonics 
of the field variables. Notice, that the scale-dependence is entirely 
encoded in the coupling constants ${\tilde J}_k, {\tilde u}_{nm}(k)$ 
and ${\tilde v}_{nm}(k)$. Inserting this ansatz into the WH-RG 
equation (\ref{WHdim}), 
\begin{eqnarray}
&&\left(2 + k\partial_k \right) \left({1\over2}{\tilde J_k}
(\phi_1 - \phi_2)^2 
+ {\tilde U}_{k}(\phi_1,\phi_2) \right) = \\ \nonumber
&&-\alpha_2 \ln\left(1+ 2{\tilde J_k}+ {\tilde U^{11}_k} + 
{\tilde U^{22}_k} + 
{\tilde J_k} ({\tilde U^{11}_k} + {\tilde U^{22}_k} + 
{\tilde U^{12}_k}) +
{\tilde U^{11}_k}{\tilde U^{22}_k} - ({\tilde U^{12}_k})^2
\right),
\end{eqnarray}
and separating the periodic and non-periodic parts, one arrives at 
the RG flow equations 
\begin{eqnarray}\label{whlsg}
\left(2 + k\partial_k \right){\tilde J_k} &=& 0, \\ \nonumber 
\left(2 + k\partial_k \right){\tilde U}_{k}(\phi_1,\phi_2) &=& 
-\alpha_2 \ln\left(1+ 2{\tilde J_k}+ {\tilde U^{11}_k} 
+ {\tilde U^{22}_k} + {\tilde J_k} ({\tilde U^{11}_k} + {\tilde U^{22}_k} 
+ {\tilde U^{12}_k}) \right. \\ \nonumber 
&&+ \left.{\tilde U^{11}_k}{\tilde U^{22}_k} - ({\tilde U^{12}_k})^2
\right)
\end{eqnarray}
since the argument of the logarithm is a periodic function 
of the fields. Notice that the dimensionless coupling constant 
$\tilde J_k$ has only a trivial tree-level evolution and hence the 
dimensionful $J = k^2 {\tilde J_k}$ remains constant during the 
RG procedure. It is also important to note that the second equation 
of (\ref{whlsg}) keeps the periodicity with the same length of 
period, therefore $\beta$ has no scale-dependence in the LPA. 

In order to specify the tree-level blocking for the generalized LSG 
model, one has to insert the ansatz (\ref{def1}) and (\ref{def2}) 
into the tree-level relation (\ref{treenew2}). For the sake of 
simplicity we turn back to dimensionful parameters and the relation 
(\ref{treenew}). Performing the integral and separating the periodic 
and the non-periodic parts, Eq. (\ref{treenew}) reduces to two 
relations. One of them,
\begin{eqnarray}
J_{k-\Delta k} = J_{k},  
\end{eqnarray}
says that the dimensionful coupling constant $J$ remains 
unchanged under the tree-level blocking, so that the 
corresponding dimensionless coupling keeps its  trivial scaling 
${\tilde J}_k = J k^{-2} $ in LPA even below the critical scale $k_c$.
The relation for the periodic part is 
\begin{eqnarray}\label{treelsg}
&& U_{k-\Delta k}(\phi_1,\phi_2) = \min_{\rho_k} 
\left\{ 2 \rho^2_k (k^2 + J) \right.  +  \\ \nonumber
&& \sum_{n,m} \big[ u_{nm}(k) \cos(n \beta \phi_1)\cos(m \beta \phi_2) 
J_0 (2 n \beta \rho_k) J_0 (2 m \beta \rho_k) \big. \\ \nonumber
&& + \big. \left. v_{nm}(k) \sin(n \beta \phi_1) \sin(m \beta \phi_2) 
J_0 (2 n \beta \rho_k) J_0 (2 m \beta \rho_k)
\big]
\right\},
\end{eqnarray}
where $J_0 (x)$ stands for the Bessel function.

Eqs. (\ref{whlsg}) and (\ref{treelsg}) can be used for the 
determination of the RG flow of the generalized LSG model in the LPA, 
but they are solvable only numerically. Instead, we give below an 
analytic solution of the linearized RG equations which is valid only 
in the UV scaling regime. Nevertheless, it has the value of being 
closely related to the result obtained in the dilute gas approximation 
\citep{pierson_lsg,pierson_rg} and with its help one can recognize
 the improvement to be expected form the numerical determination
of the WH-RG flow in the LPA.

\section{\label{lin} Linearized RG}
In order to discuss the linearized RG flow of the generalized LSG 
model, the dimensionless WH-RG equation (\ref{WHdim}) obtained in 
LPA is linearized around the Gaussian fixed point 
\begin{equation}\label{WHlin}  
\left(2 + k\partial_k \right) {\tilde V_k(\phi_1,\phi_2)} =
\,-  \, \alpha_2 \, \left({\tilde V^{11}_k} + {\tilde V^{22}_k}\right)
\end{equation}
by assuming $1 >> |{\tilde V^{ij}_k}|$ where 
${\tilde V^{ij}_k} = \partial_{\phi_i}\partial_{\phi_j} 
{\tilde V_k (\phi_1, \phi_2)}$. Using the ansatz (\ref{def1}) and  
(\ref{def2}) for the generalized LSG model, the derivatives of the 
potential are 
\begin{eqnarray}\label{linderiv} 
{\tilde V^{11}_k} &=& {\tilde J_k} - 
\sum_{n,m}  n^2 \beta^2 \big[ {\tilde u}_{nm}(k) 
\cos(n \beta \phi_1)\cos(m \beta \phi_2) 
+ {\tilde v}_{nm}(k) 
\sin(n \beta \phi_1) \sin(m \beta \phi_2) \big], \\ \nonumber
{\tilde V^{22}_k} &=& {\tilde J_k} - 
\sum_{n,m} m^2 \beta^2 \big[ {\tilde u}_{nm}(k) 
\cos(n \beta \phi_1)\cos(m \beta \phi_2) 
+ {\tilde v}_{nm}(k) 
\sin(n \beta \phi_1) \sin(m \beta \phi_2) \big].
\end{eqnarray}
Inserting Eqs. (\ref{linderiv}) into the linearized WH-RG Eq. 
(\ref{WHlin}) the following RG equations are obtained for the
coupling constants of the generalized LSG model
\begin{eqnarray}\label{lincouplings} 
(2 + k \partial_k) {\tilde J}_k &=& 0, \\ \nonumber 
(2 + k \partial_k) {\tilde u}_{nm}(k) &=& \alpha_2 \beta^2 (n^2 +m^2 )  
{\tilde u}_{nm}(k), \\ \nonumber
(2 + k \partial_k) {\tilde v}_{nm}(k) &=& \alpha_2 \beta^2 (n^2 +m^2 )  
{\tilde v}_{nm}(k) ,
\end{eqnarray}
where $\alpha_2 = 1/4 \pi$. The  linearized RG equations for the 
various coupling constants decouple and their solutions can be 
obtained analytically
\begin{eqnarray}\label{linsol} 
{\tilde J}_k &=&  J_{k=\Lambda} \left({k\over\Lambda}\right)^{-2}, 
\\ \nonumber 
{\tilde u}_{nm}(k) &=& {\tilde u}_{nm}(k=\Lambda)   
\left({k\over\Lambda}\right)^{-2 + \alpha_2 \beta^2 (n^2 +m^2 )},  
\\ \nonumber
{\tilde v}_{nm}(k) &=& {\tilde v}_{nm}(k=\Lambda)   
\left({k\over\Lambda}\right)^{-2 + \alpha_2 \beta^2 (n^2 +m^2 )},
\end{eqnarray}
where $ J_{k=\Lambda}, {\tilde u}_{nm}(k=\Lambda)$ and 
${\tilde v}_{nm}(k=\Lambda)$ are the initial values for the coupling
constants at the UV cutoff $\Lambda$. The linearized RG flow predicts 
two phases of the model. If the parameter $\beta^2$ is larger than 
a critical value $\beta^2 > \beta^2_c = 8\pi$, all the Fourier 
amplitudes ${\tilde u}_{nm}(k)$  and ${\tilde v}_{nm}(k)$ are 
irrelevant, they decrease when the cutoff $k$ is moved towards zero.
For $\beta^2 < 8\pi$, at least one of the Fourier amplitudes  
becomes  relevant. This phase structure is very similar to that 
obtained for the massless SG model \citep{sg,sg3}. Nevertheless, 
there is a significant difference. In case of the generalized LSG 
model, the coupling constant $\tilde J_k$ is always a relevant 
parameter and, consequently, the linearization  loses its validity 
with decreasing scale $k$ for any value of $\beta$. Therefore, one 
has to determine the WH-RG flow with taking all the non-linear 
terms into account in order to obtain  reliable IR behaviour 
for the coupling constants. The numerical solution of the problem is 
in progress.

Finally,  the UV scaling laws obtained above and applied to
the LSG model (\ref{lsg}) via
\begin{equation}
{\tilde u}_{10}(k) = {\tilde u}_{01}(k) = {\tilde u}(k), 
\end{equation} 
are compared to the flow given by Eq. (\ref{dilute}) found in Ref. 
\citep{pierson_lsg,pierson_rg} in the dilute gas approximation.
The linearized  WH-RG equations (\ref{lincouplings}) applied to
 the LSG model reduce to
\begin{equation}
k {d\over dk} \beta^2 = 0, 
\hspace{0.5cm}
k {d\over dk} {\tilde u} =  {\tilde u} 
\left({\beta^2\over 4 \pi} -2 \right),
\hspace{0.5cm}
k {d\over dk} {\tilde J} = - 2 {\tilde J}.
\end{equation}
These turn out to be the same as those of Eq. 
(\ref{dilute}) found in Ref. \citep{pierson_lsg} except of the loss 
of the scale-dependence of $\beta$ due to the usage of the LPA. 
Therefore, our approach with the non-linear terms included in the 
RG equations offers the possibility to achieve an improvement in the
determination of RG flow as compared to the results of the dilute gas
 approximation.

\section{\label{sum} Summary}
The new qualitative picture of the vortex length-scale dependence 
obtained by recent current transport measurements on BSCCO high 
temperature superconductor (HTSC) single crystals indicates the 
need for a better description of the critical behaviour of vortices 
in layered systems. In this paper we investigated the phase structure 
and the critical behaviour of the layered sine-Gordon (LSG) model which 
is assumed to be one of the models  describing the vortex dominated 
properties of HTSC materials reasonably well. The renormalization of 
the LSG model has been considered by the differential renormalization 
group (RG) method using a sharp momentum cutoff. The usage of sharp 
cutoff is a necessity if a  spinodal instability occurs during the 
renormalization of the model. The Wegner-Houghton (WH) RG equation as 
well as the tree-level blocking relation (which one should use in case 
of spinodal instability) have been derived in the local potential 
approximation for the LSG model. The UV scaling laws and the phase
structure  of the LSG model have been discussed at the linearized 
level by the WH-RG method. The well-known UV scaling laws
obtained in the dilute gas approximation previously in the literature
\citep{pierson_lsg,pierson_rg} were recovered. Finally, it was argued 
that our approach offers an improvement in the determination of the RG 
flow as compared to that found in the dilute gas approximation when 
all the non-linear terms are taken into account in the WH-RG equation.

\begin{acknowledgments}
I. N. thanks the Max--Planck--Institute for Nuclear 
Physics, Heidelberg, for the kind hospitality extended on the 
occasion of a guest researcher appointment in 2004 during which 
part of this work was completed. I.N. also takes a great pleasure 
in acknowledging discussion with K. Vad, S. M\'esz\'aros and 
U.D. Jentschura. This work has been supported by the grant 
OTKA T032501/00 and also partially by the grant  OTKA M041537 and  
the Supercomputing Laboratory of the Faculty of Natural Sciences, 
University of Debrecen.

\end{acknowledgments}

\end{document}